\documentclass{amsart}
\usepackage{graphicx}
\usepackage{amsmath}
\usepackage{amssymb}%
\usepackage{amsfonts}%
\usepackage{graphicx}
\usepackage{xcolor}
\usepackage{epsfig}
\usepackage[all]{xy}
\usepackage[mathscr]{euscript}
\usepackage{hyperref}
\newtheorem{theorem}{Theorem}[section]

\theoremstyle{definition}

\newtheorem{observation}[theorem]{Observation}

\usepackage[all]{xy}
\usepackage{graphicx}
\theoremstyle{remark}

\numberwithin{equation}{section}
\usepackage{graphicx}

\begin{document}

\title{The Radiant Massive Magnetic Dipole}

\author{José Diaz Polanco} \email[e-mail: ]{joseludi@unap.cl}

\address{Instituto de Ciencias Exactas y Naturales, Facultad de Ciencias, Universidad Arturo Prat, Avenida Arturo Prat 2120, Iquique, Chile.}

\author{José Ayala Hoffmann }\email[e-mail: ]{jayalhoff@gmail.com}

\address{Facultad de Ingeniería
Universidad de Tarapacá}

\author{Maximiliano Ujevic}\email[e-mail: ]{mujevic@ufabc.edu.br}

\address{Centro de Ciências Naturais e Humanas, Universidade Federal do ABC, Santo André 09210-170, SP, Brazil}
\subjclass[2000]{}

\baselineskip=20 true pt
\maketitle \baselineskip=1.10\normalbaselineskip

\maketitle

\begin{abstract} \baselineskip=20 true pt
\maketitle \baselineskip=1.10\normalbaselineskip
We present an exact, time-dependent solution for the Einstein field equations that models the coupling between an anisotropic fluid and a magnetic field in an axially symmetric space-time. By carefully selecting the metric components, we achieve a convenient separation of variables that enables us to solve Einstein's field equations and obtain a solution that evolves into the Gutsunaev-Manko massive magnetic dipole. The analysis of the thermodynamic quantities suggests that this solution may represent a pulse of radiation emitted by a massive object with magnetic properties as for example pulsars or neutron stars. 
\end{abstract}

\section{introduction}

In this note we present an exact time-dependent solution to the Einstein field equations that describes the dynamic coupling between a dipolar magnetic field and an anisotropic imperfect fluid. Our energy-momentum tensor comprises two time-dependent components, namely an anisotropic viscous fluid and a magnetic field.

Since a magnetic dipolar field has a distinguish direction we choose an axially symmetric time-dependent metric. In addition, since we are going to include a fluid under this setup it is reasonable to consider the existence of an anisotropy (in the thermodynamics variables). On the other hand, note that the magnetic field may produce an viscous effect on the fluid. We strategically consider a time-dependent metric in such a way that the Einstein field equations can be written as a system of separable differential equations.

Under the time evolution of the metric functions of our model we see that the final state of equilibrium is equivalent to the Gutsunaev and Manko metric implying that our solution includes the one of Gutsunaev and Manko \cite{gus} as a particular case as $t\to\infty$. This result is coherent after noticing that the equilibrium state is reached when the dissipative variables disappear. For this reason, and according to \cite{karas}, we infer that our model leads to a central massive body exhibiting the characteristics of a magnetic dipole. We call this solution {\bf Massive Radiant Magnetic Dipole} (MRMD in short). By setting the current density to zero, we find that the Ricci scalar is also null, and the effective pressure is $1/3$ of the energy density, indicating a radiation-like fluid, see Equation (\ref{eq:PRESTOTAL1}). The analysis of thermodynamic quantities suggests that our solution could represent a magnetized large object emitting radiation that decays rapidly in time.

In a previous study \cite{pepe}, we study the case of a massive magnetic dipole surrounded by a charged static fluid. Therein we presented a stationary solution that couples an electromagnetic field, a fluid, and gravity. Later, Cabrera-Munguia and Manko extended this solution by showing the existence of a family of non-trivial solutions in which the coupling between the magnetic field and a specific fluid is successful. However, note that this solution is limited to the static case, see \cite{cmanko}.

The main contributions of this article are:
\begin{enumerate}
    \item We found an exact solution to the Einstein field equations that accounts for the coupling between an anisotropic fluid and a magnetic field. This solution is time-dependent and considers the space-time to be axially symmetric. In addition, all the thermodynamic variables are calculated respecting causality. 
    
    \item The physical interpretion of our model is allingned with several astrophysical systems with magnetic fields such pulsars or neutron stars. Also, the physical interpration of our model is allingned with several astrophysical systems with magnetic fields such as neutron stars or pulsars.

    \item We found an interesting relation between the temperature and the magnetic dipole moment in which we conjecture that this could be related with the temperature anomaly at the solar corona, see Figure \ref{fig:pintinho}.
\end{enumerate}

This article is structured as follows. In Section \ref{two}, we introduce the metric and energy-momentum tensor that we will be considering. Then, in Section \ref{three}, we present the solution for the Einstein field equations for this model. Moving on, in Section \ref{four}, we investigate the thermodynamic properties of the radiation that is coupled with the magnetic field and analyze the time-dependent energy fluxes. In Section \ref{five}, we examine the temporal evolution of the energy fluxes. Lastly, in Section \ref{six}, we summarize our findings.

\section{coupling field for the proposed model} \label{two}

Our goal is to model the coupling between an anisotropic imperfect fluid and a dipolar magnetic field. This requires solving the Einstein field equations, which describe the fundamental relationship between the geometry of spacetime and the distribution of matter and energy within it given by,
\begin{equation}
G_{\alpha \beta }=8\pi \left(\chi_{\alpha \beta }+\Pi_{\alpha \beta
}\right)
\end{equation}
here $G_{\alpha \beta }$ is the Einsten tensor, $\chi_{\alpha \beta }$ is the electromagnetic tensor and
$\Pi_{\alpha \beta }$ is the fluid tensor. As costumary, we use greek indices from 1 to 4, and latin indices running from 1 to 3.
We use the cylindrical Weyl–Lewis–Papapetrou coordinates in $3 + 1$ spacetime, so that $(x^{1},x^{2},x^{3},x^{4})=(\rho ,z,\varphi ,t)$.

The spacetime for our model is represented by the special Weyl metric,
\begin{equation}
ds^{2}=\frac{a}{\mathcal{F}}\left( e^{2\mathcal{H}
}\left[ d\rho ^{2}+dz^{2}\right] +\rho ^{2}d\varphi ^{2}\right)
-\frac{\mathcal{F}}{a}dt^{2}
 \mbox{}\label{eq:ds2t}
\end{equation}
here $\mathcal{F}, \mathcal{H}: \mathbb R^2 \to \mathbb R$ are functions of the variables $\rho, z$, and $a$ is a single real valued function of $t$. 
 
We consider $G=c=1$. Also the partial and
covariant derivatives with respect to the coordinate $x_{\mu }$ are denoted by ,$\mu $ and ;$\mu$ respectively. 

The electromagnetic energy-momentum tensor,
\begin{equation*}
\chi_{\alpha \beta }=\frac{1}{4\pi }\left[ F_{\alpha \tau }\ F_{\beta
}^{\text{ \ }\tau }-\frac{1}{4}\ \ g_{\alpha \beta }F_{\tau \sigma
}\ F_{\text{ \ \ }}^{\text{ }\tau \sigma }\right]
\end{equation*}
is defined appropriately through the 4-potential,
\begin{equation}
A_{\mu }=\left[0,0,\sqrt{a\left(t\right)} \mathcal{A},0
\right]\label{eq:AAA} 
\end{equation}
where $\mathcal{A}$ is also a function of $(\rho ,z)$.

To define the energy-momentum tensor associated with an anisotropic fluid we use the definition of a tetrad basis
$\left(e_{(\gamma)}^{\alpha} \right)$ so that,
\begin{equation}
g_{\alpha\beta}e_{(\gamma)}^{\alpha}e_{(\lambda)}^{\beta}=\eta_{(\gamma)(\Lambda)}
\end{equation}
where  $\eta_{(j)(k)}$ corresponds to the Minkowski metric. We can choose the following tetrad basis:
\begin{eqnarray}
e_{(1)}^{\alpha} &=&\left[e_{(1)}^{1},e_{(1)}^{2},e_{(1)}^{3},e_{(1)}^{4}\right]=\left[\frac{1}{\sqrt{g_{11}}},0,0,0\right]\\
e_{(2)}^{\alpha} &=&\left[e_{(2)}^{1},e_{(2)}^{2},e_{(2)}^{3},e_{(2)}^{4}\right]=\left[0,\frac{1}{\sqrt{g_{22}}},0,0\right]\\
e_{(3)}^{\alpha} &=&\left[e_{(3)}^{1},e_{(3)}^{2},e_{(3)}^{3},e_{(3)}^{4}\right]=\left[0,0,\frac{1}{\sqrt{g_{33}}},0\right]\\
e_{(4)}^{\alpha}
&=&\left[e_{(4)}^{1},e_{(4)}^{2},e_{(4)}^{3},e_{(4)}^{4}\right]=\left[0,0,0,\frac{-1}{\sqrt{-g_{44}}}\right]\mbox{}
\end{eqnarray}
Thus the energy-momentum tensor ($\Pi_{\alpha \beta }$) may be
written in the generic form,
\begin{equation}
\Pi^{\alpha \beta }=\varepsilon
e_{(4)}^{\alpha}e_{(4)}^{\beta}+P\left(
e_{(1)}^{\alpha}e_{(1)}^{\beta}+ e_{(2)}^{\alpha}e_{(2)}^{\beta}
\right) +P_{3} e_{(3)}^{\alpha}e_{(3)}^{\beta}+\Omega^{\alpha\beta}
\label{eq:TTETR}
\end{equation}
where $\varepsilon$ is the fluid energy density, $P$ and $P_{3}$ are
the fluid pressures and $\Omega^{\alpha\beta}$ represents a stress
tensor. To obtain the elements of $\chi^{\alpha \beta }$ we use the identities
\begin{equation}
g^{\alpha\beta}=e_{(1)}^{\alpha}e_{(1)}^{\beta}+e_{(2)}^{\alpha}e_{(2)}^{\beta}
+e_{(3)}^{\alpha}e_{(3)}^{\beta}-e_{(4)}^{\alpha}e_{(4)}^{\beta}
\label{eq:STETR}
\end{equation}
\begin{equation}
 e_{(3)}^{\alpha}e_{(3)}^{\beta}=g^{33}\delta_{(3)}^{\alpha} \delta_{(3)}^{\beta}\label{eq:TR3}
\end{equation}
and furthermore, we define the 4-velocity as follows,
\begin{equation}
u^{\alpha}=e_{(4)}^{\alpha}\label{eq:4velT}
\end{equation}
Substituting (\ref{eq:STETR}-\ref{eq:4velT}) in (\ref{eq:TTETR})
and using the Eckart's conditions for an imperfect fluid (see \cite{eck}) we can write the anisotropic energy-momentum tensor in the form
\begin{equation}
\chi_{\alpha \beta }=\varepsilon u^{\alpha}u^{\beta}+\left(
 P-\varsigma \theta \right) h^{\alpha\beta} +\left(P_{3}-P
\right)g^{33}\delta_{(3)}^{\alpha} \delta_{(3)}^{\beta}+
\tau^{\alpha\beta}\label{eq:TTETRFIN}
\end{equation}
where $\varsigma $ is the bulk effective viscosity,
$\theta=u^{\alpha}_{ \ ; \alpha}$ is the expansion
$h^{\alpha\beta}=u^{\alpha}u^{\beta}+g^{\alpha\beta}$ and
$\tau_{\alpha \beta }$ is,
\begin{equation}
\tau_{\alpha \beta }=-2\eta \sigma _{\alpha \beta }+q_{\alpha
}u_{\beta }+q_{\beta }u_{\alpha }
\end{equation}
Note that $q^{\alpha }$ is the heat flux, $\eta $ is the shear
viscosity and $\sigma _{\alpha \beta }$ is the symmetric trace-free
spatial shear tensor given by,
\begin{equation}
\sigma _{\alpha \beta }=\frac{1}{2}\left\{ (u_{\alpha }h_{\text{ \
}\beta }^{\mu })_{;\mu }+(u_{\beta }h_{\text{ \ }\alpha }^{\mu
})_{;\mu }\right\} -\frac{1}{3}\theta h_{\alpha \beta }
\end{equation}
In this model the 4-velocity, the energy flux density, fluxes, and
pressures are also time dependent, that is to say,
\begin{eqnarray*}
u^{4}&=&\frac{-\sqrt{a\left(t\right)}}{\sqrt{\mathcal{F}}}\hspace{0.6cm}\varepsilon=\varepsilon\left(\rho,z,t\right)
\hspace{0.6cm}P=P\left(\rho,z,t\right)
 \\ P_{3}&=&P_{3}\left(\rho,z,t\right)
\hspace{0.6cm}q_{\alpha}=q_{\alpha}\left(\rho,z,t\right)
\end{eqnarray*}
With these considerations, in the next section we present explicitly the Einstein field equations associated with our model.

\section{einstein field equations for the proposed model} \label{three}

The Einstein field equations for the coupling of an anisotropic imperfect fluid with a magnetic field is given by the following a set of equations:

\begin{eqnarray}
\mathcal{H} _{,\rho }&=&\frac{\rho}{4}\frac{ \mathcal{F}_{,\rho
}^{2}-\mathcal{F}_{,z}^{2}}{\mathcal{F}^{2}}+\frac{\mathcal{F}}{\rho}\left(
\mathcal{A}_{,\rho }^{2}
-\mathcal{A}_{,z}^{2}\right)  \label{eq:EGUS111}\\
\mathcal{H} _{,z}&=&\frac{\rho}{2}\frac{
\mathcal{F}_{,z}\mathcal{F}_{,\rho
}}{\mathcal{F}^{2}}+\frac{2\mathcal{F}}{\rho
}\mathcal{A}_{,\rho }\ \mathcal{A}_{,z} ,\label{eq:EGUS211}\\
\mathcal{H} _{,z,z}+\mathcal{H} _{,\rho ,\rho
}&=&-\frac{1}{4}\frac{\mathcal{F}_{,\rho
}^{2}+\mathcal{F}_{,z}^{2}}{\mathcal{F}^{2}}+\frac{\mathcal{F}}{\rho
^{2}}\left( \mathcal{A}_{,\rho }^{2}
+\mathcal{A}_{,z}^{2}\right)  \label{eq:EGUS311}\\
\varepsilon&=&\frac{\left(3 \rho ^2-\mathcal{F}\mathcal{A}^2\right) a'(t)^2}{32 \pi  \rho ^2 a(t) \mathcal{F}}\\
P+\frac{a''(t)}{8 \pi \mathcal{F}}&=&\frac{(-3 \zeta +2 \eta )
a'(t)}{2 \sqrt{a(t)} \sqrt{\mathcal{F}}}-\frac{\left(\rho
^2+\mathcal{F} \mathcal{A}^2\right) a'(t)^2}{32 \pi \rho ^2
a(t)\mathcal{F}} \label{eq:EGUSP}\\
P_{3}+\frac{a''(t)}{8 \pi \mathcal{F}}&=&\frac{(-3 \zeta +2 \eta )
a'(t)}{2 \sqrt{a(t)} \sqrt{\mathcal{F}}}-\frac{\left(\rho
^2-\mathcal{F} \mathcal{A}^2\right) a'(t)^2}{32 \pi \rho ^2
a(t)\mathcal{F}} \label{eq:EGUSP3}\\
q_{1}&=&\frac{a'(t) \left(\rho ^2 \mathcal{F}_{\rho}-2 \mathcal{F}^2
\mathcal{A} \mathcal{A}_{\rho}\right)}{16 \pi  \rho ^2 \sqrt{a(t)}
\mathcal{F}^{3/2}}  \label{eq:EQ1}\\
q_{2}&=&\frac{a'(t) \left(\rho ^2 \mathcal{F}_{z}-2 \mathcal{F}^2
\mathcal{A} \mathcal{A}_{z}\right)}{16 \pi  \rho ^2 \sqrt{a(t)}
\mathcal{F}^{3/2}} \label{eq:EQ2} \\
q_{3}&=&0.\label{eq:EQ3}
\end{eqnarray}
Note that (\ref{eq:EGUS111}-\ref{eq:EGUS311}) can be solved using the integrability condition $\mathcal{H}_{,\rho ,z}=\mathcal{H}_{,z,\rho }$ and the substitution of (\ref{eq:EGUS111}-\ref{eq:EGUS211}) in (\ref{eq:EGUS311}). Both conditions can be written as \ref{eq:condition1} and \ref{eq:condition2}. Even more, we can remove
$\mathcal{H}$ in order to obtain the system equations in terms of
$\mathcal{F}$ and $\mathcal{A}$ as follows,
\begin{eqnarray}
\overrightarrow{\nabla }\cdot \left( \frac{\rho }{\mathcal{F}
}\overrightarrow{\nabla }\mathcal{F}\right)
-\frac{2\mathcal{F}}{\rho }\overrightarrow{\nabla } \mathcal{A}\cdot
\overrightarrow{\nabla }\mathcal{A} =0\label{eq:condition1}\\
\overrightarrow{\nabla }\cdot \left(
\frac{\mathcal{F}}{\rho}\overrightarrow{\nabla
}\mathcal{A}\right)=0 \label{eq:condition2}
\end{eqnarray}
where $\overrightarrow{\nabla }=\hat{\rho} \partial_{\rho}+\hat{z} \partial_{z}$ is the 2-dimensional nabla operator in the $\left( \rho, z \right)$ coordinates. 

For simplicity, equations (\ref{eq:condition1}) and (\ref{eq:condition2}) can be written in prolate ellipsoidal coordinates $x$ and $y$ defined by, 

\begin{equation}
\rho =k\sqrt{x^{2}-1}\sqrt{1-y^{2}} \hspace{0.5cm}   z=kxy  
\end{equation}
where $k$ is an appropiate scale factor. In these news coordinates there is a solution for $\ref{eq:condition1}$ and $\ref{eq:condition2}$ which is given in the form:

\begin{eqnarray}
\mathcal{F}&=&\frac{x-1}{x+1}\left(\frac{\left[x^{2}-y^{2}+\alpha^{2}(x^{2}-1)\right]^{2}
+4 \alpha^{2}
x^{2}(1-y^{2})}{\left[x^{2}-y^{2}+\alpha^{2}(x-1)^{2}\right]^{2} -4
\alpha^{2}y^{2}(x^{2}-1)} \right)^{2} \label{eq:solF} \\
\mathcal{A}&=&\frac{4k
  \alpha^{3}\left(1-y^{2}\right)\left[2\left(
\alpha^{2}+1\right)x^{3}+\left(1-3\alpha^{2}\right)x^{2}+y^{2}+\alpha^{2}\right]}
{\left(\alpha^{2}+1\right) \left( \left[x^{2}-y^{2}+\alpha^{2}
\left(x^{2}-1 \right) \right]^{2}+4\alpha^{2} x^{2}
\left(1-y^{2}\right) \right)} \label{eq:solA} \\
e^{2\mathcal{H}}&=&\frac{x^{2}-1}{x^{2}-y^{2}}\frac{\left(
\left[x^{2}-y^{2}+\alpha^{2}(x^{2}-1)\right]^{2}+4
\alpha^{2}x^{2}(1-y^{2})\right)^{4}}
{(\alpha^{2}+1)^{8}(x^{2}-y^{2})^8} \label{eq:solH}
\end{eqnarray}

We choose $k$ such that the solution of equation \ref{eq:ds2t} is asimptotically equivalent to Schwarzchild's metric.
If, 
\begin{eqnarray}
x=\frac{1}{k} \left(r-m\right)\,\,\, \mbox{and}\,\,\, y=\cos \theta \mbox{
}\label{eq:COORD}
\end{eqnarray}
where,
\begin{equation}
k=\frac{m\left(1+\alpha^{2}\right)}{\left(1-3\alpha^{2}\right)}
\end{equation}
then, 
\begin{equation}
\mathcal{F}=1-\frac{2m}{r}+O\left(
r^{-3}\right)
\end{equation}
and,
\begin{equation}
e^{2\mathcal{H}}=1-O\left(r^{-2}\right)
\end{equation}
Similarly, by expanding the potential function $\mathcal{A}$ in terms
of $r^{-1}$, we can write it as follows,
\begin{equation}
\mathcal{A}=\frac{8m^{2}\alpha^{3}\sin^{2}\theta}{\left(1-3\alpha^{2}\right)^{2}r}+\frac{12m^{3}\alpha^{3}\sin^{2}\theta}{\left(1-3\alpha^{2}\right)^{2}r^{2}}+O\left(r^{-3}\right) \mbox{}
\end{equation}
Note that the asymptotic form for the potential
electromagnetic field is of the magnetic dipole type. In this fashion, we can define the magnetic moment $\mu$ as follows,
\begin{equation}
\mu=\frac{8m^{2}\alpha^{3}}{\left(1-3\alpha^{2}\right)^{2}} \mbox{}\label{eq:MMU}
\end{equation}
implying that,
\begin{equation}
\mathcal{A}=\frac{\mu \sin^{2}\theta}{r}+O\left(r^{-2}\right)\mbox{}
\end{equation}
Note that for $\alpha=0$ we recover the solution of
Schwarzschild.

By considering functions $\mathcal{F}, \mathcal{A}$ and
$\mathcal{H}$ given in (\ref{eq:solF}-\ref{eq:solH}) we can explicitly calculate the 4-current components of $J_{\alpha}$ to obtain,
\begin{eqnarray}
J^{1}&=&0\\
J^{2}&=&0\\
J^{3}&=&-\frac{\mathcal{A}(a'(t)^{2}+2a(t)a''(t))}{16 \pi \rho^{2}
a(t)^{3/2}}\\
J^{4}&=&0
\end{eqnarray}
implying that the current density (according to $J^{4}=\rho
v^{4}$) is null. This impose a condition for the temporal
function $a(t)$ which can be written in the form,
\begin{equation}
a'(t)^{2}+2a(t)a''(t)= 0 \label{eq:eqdifa}
\end{equation}

Condition \ref{eq:eqdifa} together with the temporal function $a(t)$, provide us information about the Ricci scalar $R$. We can calculate explicitly the Ricci scalar to obtain,
\begin{equation}
R=\frac{3(a'(t)^{2}+2a(t)a''(t))}{2 a(t) \mathcal{F}}
\end{equation}
and, according to (\ref{eq:eqdifa}), the Ricci
scalar should be null. The solution of equation
(\ref{eq:eqdifa}) may be written in the form,
\begin{equation}
a(t)=a_{0}\sqrt[3]{\left(\frac{3}{2}(a_{1}t+1)\right)^2}
\label{eq:at}
\end{equation}
where $a_{0}$  and $a_{1}$ are constant. With this result at hand we can
guarantee that all the system thermodynamic properties can be calculated explicitly.

\section{the thermodinamic properties of the system} \label{four}

In this section we explicitly calculate the thermodynamic properties of the system in terms of $\mathcal{F}$, $\mathcal{A}$, $\mathcal{H}$, and $a(t)$, all of which are known. Our first step is to calculate the temperature of the system. To this end, we calculate the system energy total fluxes according to Eckarts' thermodynamic which is refering from now on as,
\begin{equation}
Q_{\alpha }=-\kappa h_{\alpha}^{\beta } \left[ T_{,\beta}+Tu^{\mu }v_{\beta ; \mu}\right]\label{eq:QQQ}
\end{equation}
where $\kappa$ is the thermal conductivity and $T$ is the
temperature of the system. We have explicitly that,
\begin{eqnarray}
Q_{1 }&=&-\kappa  \left(\frac{T \mathcal{F}_{\rho}}{2 \mathcal{F}}+T_{\rho}\right)\label{eq:Q11}\\
Q_{2 }&=&-\kappa  \left(\frac{T \mathcal{F}_{z}}{2
\mathcal{F}}+T_{z}\right)\label{eq:Q22}\\
Q_{3 }&=&0
\end{eqnarray}
On the other hand, the resulting flux $Q_{\alpha }$ is directly calculated as the spatial component of the  4-flux of energy,
\begin{equation}
Q_{\alpha }=T_{\alpha}^{\beta } v_{\beta } \label{eq:QT}
\end{equation}
where $T_{\alpha \beta }$ is the total energy-momentum tensor. Thus,
using (\ref{eq:QT}) we obtain,
\begin{eqnarray}
Q_{1 }&=&q_{1 }+\frac{\sqrt{\mathcal{F}} \mathcal{A} a'(t)
\mathcal{A}_{\rho}}{8 \pi \rho ^2 \sqrt{a(t)}}=\frac{a'(t) \mathcal{F}_{\rho}}{16 \pi  \sqrt{a(t)} \mathcal{F}^\frac{3}{2}}\label{eq:Q111}\\
Q_{2 }&=&q_{2 }+\frac{\sqrt{\mathcal{F}} \mathcal{A} a'(t)
\mathcal{A}_{z}}{8 \pi \rho ^2 \sqrt{a(t)}}=\frac{a'(t)
\mathcal{F}_{z}}{16 \pi  \sqrt{a(t)} \mathcal{F}^\frac{3}{2}} \label{eq:Q222}\\
Q_{3 }&=&q_{3 }=0
\end{eqnarray}
where $q_{\alpha }$ corresponds to the fluxes obtained by Einstein's
equations (\ref{eq:EQ1}-\ref{eq:EQ3}). Finally, by comparing
(\ref{eq:Q11}) with (\ref{eq:Q22}) and (\ref{eq:Q111}) with
(\ref{eq:Q222}) we have the following differential equations:
\begin{eqnarray}
\frac{a'(t) \mathcal{F}_{\rho}}{16 \pi  \sqrt{a(t)} \mathcal{F}^\frac{3}{2}}&=&-\kappa
\left(\frac{T \mathcal{F}_{\rho}}{2 \mathcal{F}}+T_{\rho}\right)\label{eq:TT11}\\
\frac{a'(t) \mathcal{F}_{z}}{16 \pi  \sqrt{a(t)}
\mathcal{F}^\frac{3}{2}}&=&-\kappa  \left(\frac{T \mathcal{F}_{z}}{2
\mathcal{F}}+T_{z}\right)\label{eq:TT22} 
\end{eqnarray}
By solving these equations we can see that the temperature field of the
system is,
\begin{equation}
T\left(\rho,z,t\right)=\frac{a'(t)}{16 \pi \kappa
\sqrt{a(t)}}\frac{T_{\infty}-\ln{\mathcal{F}}}{\sqrt{\mathcal{F}}}\label{eq:TEMPER}
\end{equation}
Also, this can be used to show that the total energy $\xi$ and the
effective pressure $P_{ef}$  may be written as,
\begin{equation}
\xi\left(\rho,z,t\right)=\frac{ 3\rho ^2 a'(t)^2+4 \mathcal{F}^3
e^{-2\mathcal{H}}\left(\mathcal{A}_{\rho}^2+\mathcal{A}_{z}^2\right)}{32
\pi  \rho ^2 a(t) \mathcal{F}} \label{eq:ENERTOTAL}
\end{equation}
\begin{equation}
P_{ef}\left(\rho,z,t\right)=\frac{\xi\left(\rho,z,t\right)}{3} \label{eq:PRESTOTAL1}
\end{equation}

The relation we have found between the effective pressure $P_{ef}$ and the total energy of the system $\xi$ indicates that the effective pressure behaves like radiation pressure. Additionally, we have deduced that the Ricci scalar is null and there is no current density. From these results, we can conclude that the fluid in the system is most likely a radiation. Moreover, since the system's metric is equivalent to the Gutsunaev and Manko metric, we can also infer that there exists a central massive body exhibiting the characteristics of a magnetic dipole. We refer to this solution as the {\bf Massive Radiant Magnetic Dipole} (MRMD in short form). With the mathematical description of the model now complete, we know all the physical quantities associated with the system. Notably, we have observed that the temporal function is not oscillatory, leading us to speculate that gravitational waves are absent.

It is worth noticing the relationahip between the system pressures and how these relate with the MRMD solutions. The equations (\ref{eq:EGUSP3}) and (\ref{eq:EGUSP}) define the pressures $P$ and $P_{3}$, respectively, and both include the kinetic and dynamic viscosity coefficients. Interestingly, the term $2 \eta-3 \zeta$ is present in both pressures, suggesting that we can classify the MRMD solutions based on this factor.

The following types of MRMD solutions can be classified based on the presence and values of the viscosity pressure terms in equations (\ref{eq:EGUSP3}) and (\ref{eq:EGUSP}).

\begin{itemize}
\item Type I: No viscosity pressure terms are present, i.e., $\zeta = \eta = 0$. This case describes a radiation in the traditional context of first-order thermodynamics.
\item Type II: Viscosity pressure terms are present, but the viscosities are self-compensated, meaning that $\zeta = 2\eta/3$.
\item Type III: Viscosity pressure terms are present and the viscosities are not self-compensated; meaning that $\zeta \neq 2\eta/3$ and $\eta \neq 0$. In this case, the radiation exhibits a first-order viscosity characteristic (see \cite{is2}).
\end{itemize}

Thus, the MRMD solution depends on the type of radiation in
question. However, in the three cases previously mentioned the
physical quantities associated to the metric behave as they were
independent of the viscosity.

\section{graphical analysis for the thermodinamical variables} \label{five}

\begin{figure}[h]
\centering{\epsfig{width=12cm, height=4.5cm,
file=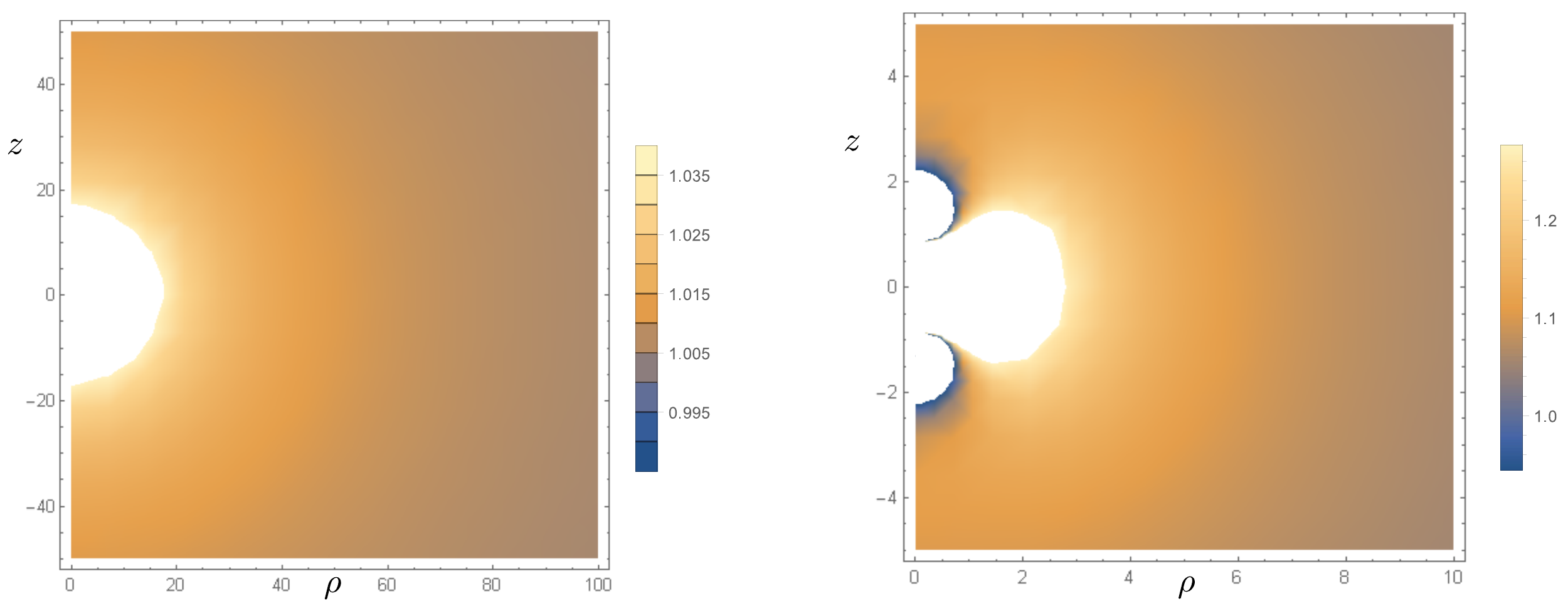}} \caption{Graphs for temperature of the fluid according to equation (\ref{eq:TEMPER}). Left: a panoramic version of the temperature graph. Right: a zoom of the temperature graph near the origin.}
\label{fig:Trad}
\end{figure}

We begin with a graphical analysis for the themperature of the  fluid according to equation (\ref{eq:TEMPER}), see Figure \ref{fig:Trad}.

 The system's temperature presents the typical behaviour of decreasing proportionally to the distance, reaching an asitontic value of $T_\infty$, which could be different from zero. This could even be adjusted to cosmological values. In contrast, the temperature near the magnetic poles is quite high.

\begin{figure}[h]
\centering{\epsfig{width=12cm, height=4.5cm,
file=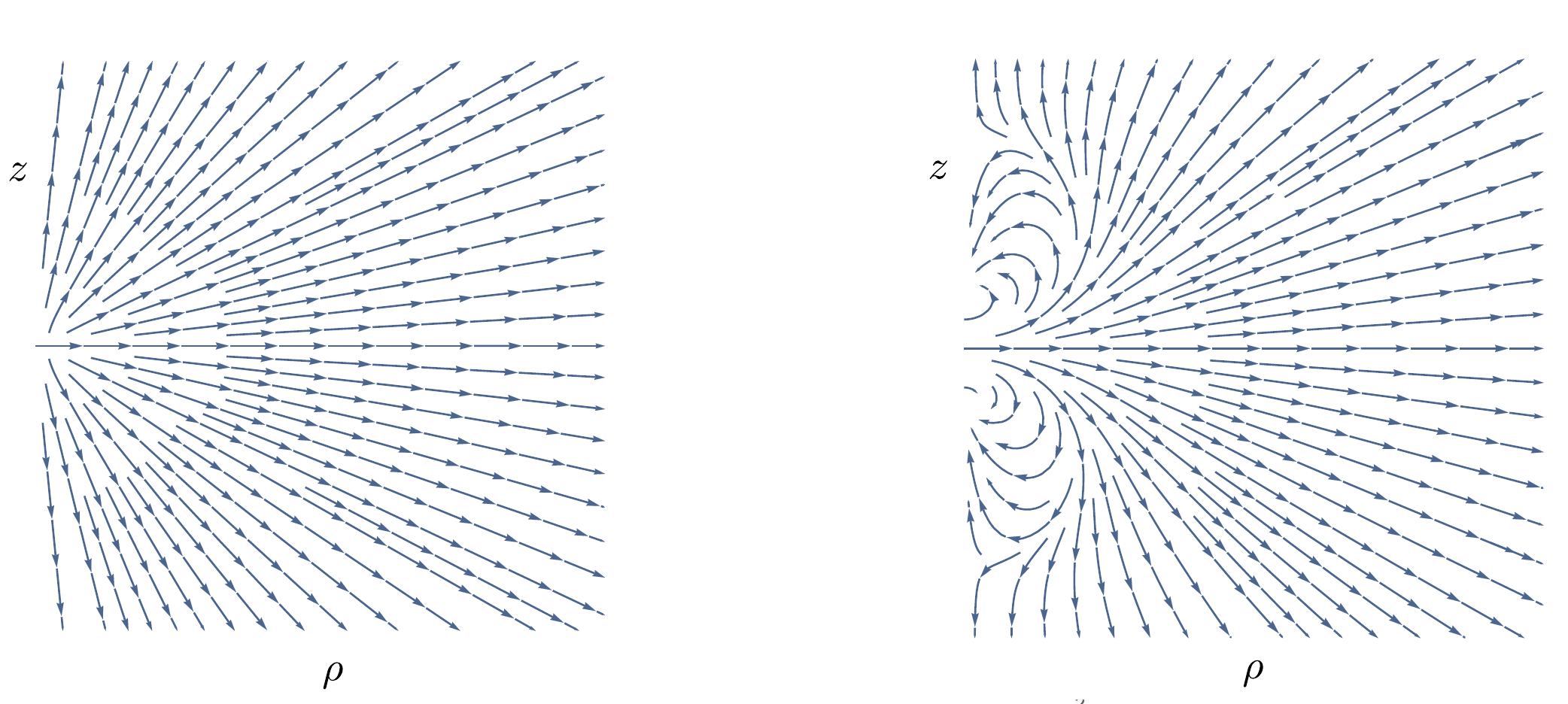}} \caption{This solution presents a radiation emitted by a central body with magnetic dipole characteristics. The right illustration represents a zoom at the origin of the figure at the left.}
\label{fig:Trad1}
\end{figure}

In order to analyse the behaviour of the total energy fluxes $Q_{\alpha}$ during the emission (see Figure \ref{fig:Trad1}) the vector field energy flux generated by $\left[Q_{1}, Q_{2}, 0\right]$ is visualized acording to equations (\ref{eq:Q111}) and (\ref{eq:Q222}). Note that the total energy flux of the sistem is composed by the heat flux and an electromagnetic flux.

Next we show a graphical reperesentation for the energy flux. This can be represented in a generic way at any given moment, see Figures \ref{fig:Trad1}.

\begin{figure}[h]\centering{\epsfig{width=7cm, height=6cm,
file=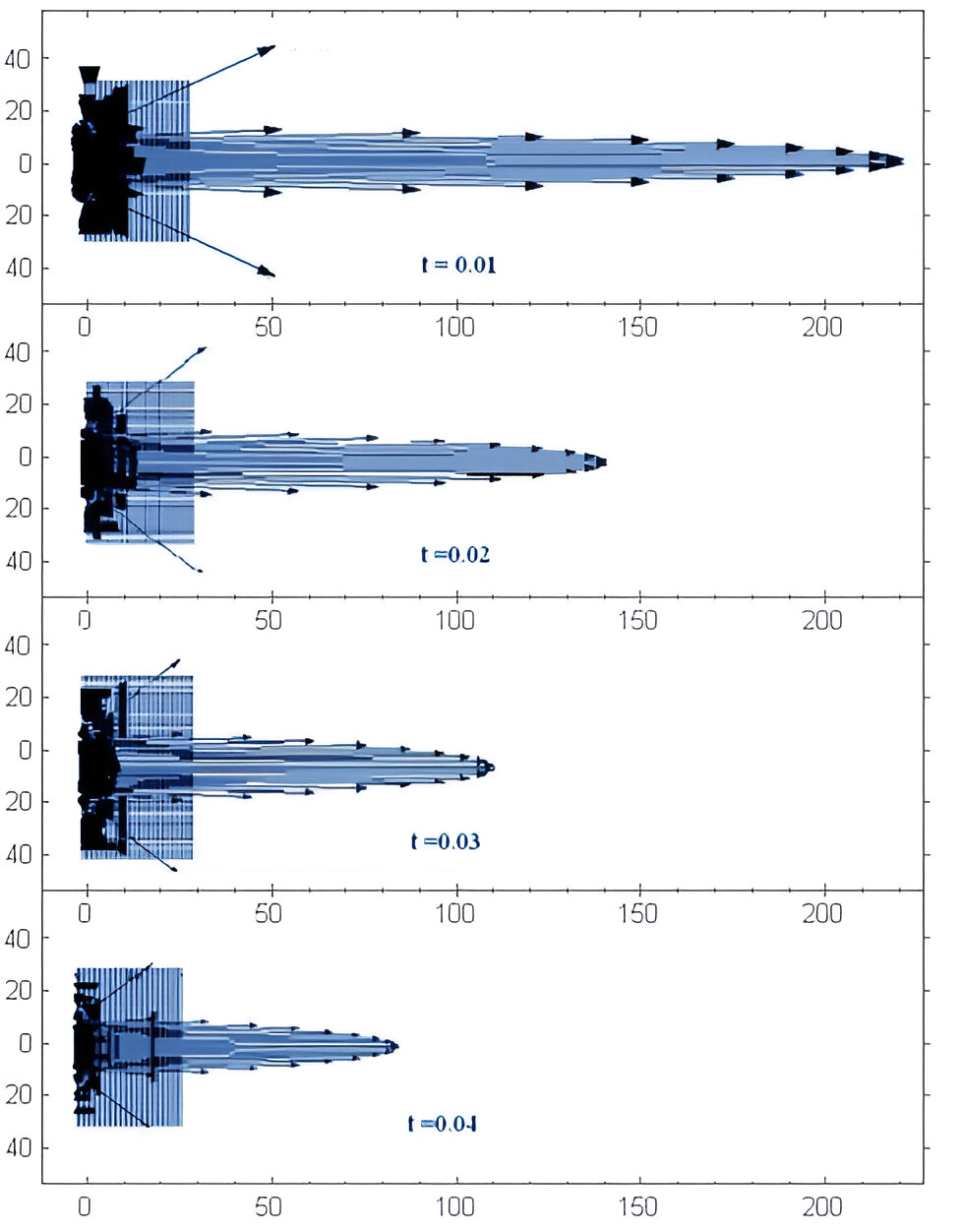}}
\caption{Evolution of the energy flux vector field for small times. Note that the energy flux intensity decreases quickly in the course of time.} \label{fig:explosion}
\end{figure}

\begin{figure}[h]
\centering{\epsfig{width=12cm, height=3.5cm,
file=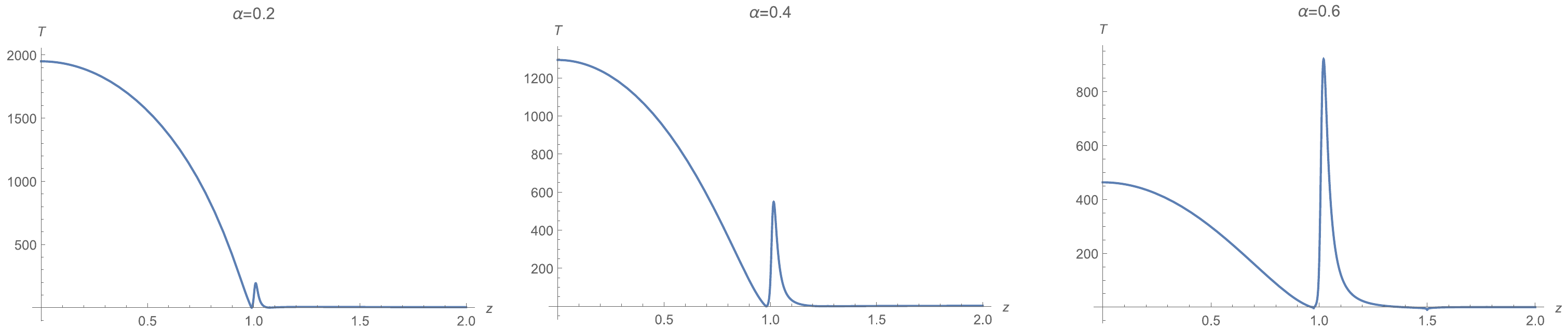}} \caption{A plot of the temperature over the $z$-axis (i.e., $\rho=0$) according to equation \ref{eq:TEMPER} for $\alpha=0.2,0.4,0.6$ at arbitrary time. Note that the temperature at $z=0$ (stellar nucleus) is inversely proportional to the magnetic parameter $\alpha$. On the other hand, the pick of temperature around the stellar corona is  directly proposional to $\alpha$. Also, for $\alpha=0$ the pick dissapears.}
\label{fig:pintinho}
\end{figure}

\newpage
\begin{observation}[Energy Flux Analysis] \hfill
\begin{itemize}

\item Note that a large energy emission occurs in the $\rho$-axis direction, as shown in Figure \ref{fig:explosion}. There are also two other distinguished directions with large fluxes.

    \item The Figure \ref{fig:explosion} illustrates that the flux intensity declines quickly, and it can be shown that the flux decreases hyperbolically regarding time, proportionally with $t^{-2/3}$. 

    \item The imperfect fluid in our model can be interpreted as a radiation pulse.

\end{itemize}

It is important to note that we have modeled a
radiation pulse explicitly using the formalism of general relativity. We think that this model is a good representation of the massive energy emanation that
befalls in astrophysical systems with magnetic fields such as pulsars, neutron stars, or other similar configurations. More complex cases like Kerr models with rotation are topics for current research.
\end{observation}

\section{conclusions and conjectures} \label{six}

We have developed an explicit time-dependent solution to Einstein's equation that couples a dipole magnetic field with a fluid that exhibits radiation properties. The spatial metric component corresponds to the metric of Gutsunaev and Manko, which describes a massive magnetic dipole in a vacuum. As time progresses, the behavior of the space-time exhibits a non-adiabatic collapse, and the emitted energy fluxes decay rapidly, much like astrophysical pulses.

Furthermore, we have identified three types of solutions that govern the radiation in the vicinity of the magnetic source, each of which is dependent on viscosity. Namely, Type I in which there are no viscosity pressure terms, which occurs when $\zeta = 2\eta /3$. Type II, in which there are also no viscosity pressure terms, which occurs when $\zeta = \eta = 0$. Type III, in which there are viscosity pressure terms, which occurs when $\zeta \neq 2\eta /3$, $\eta \neq 0$.

It is noteworthy that the space-time is independent of the viscous fluid. This model could describe a pulsar observed by an observer co-moving with the pulsar's rotation or simpler models like highly magnetized stars that emit radiation. Depending on the system's initial density and associated energy fluxes, this solution can be tailored to several astrophysical systems. This solution serves as a starting point for future research aiming to study real astrophysical objects where the interaction between matter and energy cannot be ignored. In addition, we expect that this theoretical result would be constrasted with data obtained from the solar corone. We conjecture that the temperature pick (maximum value, see Figure \ref{fig:pintinho}) would give a non trivial connection between the nucleous temperature, and the dipolar magnetic moment.


\bibliographystyle{amsplain}

\begin{thebibliography}{11} \rm
   
\bibitem{pepe} J. D. Polanco, P. S. Letelier and M. Ujevic, Phys. Rev. D {\bf 70}, 104003 (2004).

 \bibitem{cmanko} J. L. Cabrera-Munguia and V. S. Manko,  Phys. Rev. D {\bf 82}, 124042 (2010).
   
\bibitem{gus} Ts. I. Gutsunaev and V. S. Manko, Phys. Lett. A {\bf 123}, 215
(1987).

\bibitem{karas} D. Vokrouhlický, V. Karas.  Gen Relat Gravit {\bf 22}, 1033–1043 (1990)

\bibitem{eck} C. Eckart, Phys. Rev. {\bf 58}, 919 (1940).


\bibitem{is2} W. Israel and J. M. Stewart, Ann. Phys. {\bf 118}, 341
(1979).

\bibitem{nashed}  G.G.L. Nashed, S. Capozziello.  Gen Relativ Gravit {\bf 51}, 50 (2019).

\end{thebibliography}
   
\end{document}